\documentclass[10pt,twocolumn,letterpaper]{article}

\usepackage[ruled,noend]{algorithm2e} %

\SetAlFnt{\small}
\SetAlCapFnt{\small}
\SetAlCapNameFnt{\small}
\SetAlCapHSkip{0pt}
\IncMargin{-\parindent}
\usepackage{ifpdf}
\usepackage{graphicx}
\ifpdf
  
\else
\fi
\usepackage{color}
\definecolor{cyan}{cmyk}{1,0,0,0}
\definecolor{darkgreen}{rgb}{0,0.5,0}
\definecolor{orange}{rgb}{1,0.5,0}
\definecolor{magenta}{cmyk}{0,1,0,0}
\definecolor{darkyellow}{cmyk}{0,0,0.75,0}
\definecolor{gray}{rgb}{0.8,0.8,0.8}

\newenvironment{tightitemize}{
\vspace{-1.5mm}
\begin{itemize}
  \setlength{\itemsep}{1pt}
  \setlength{\parskip}{2pt}
  \setlength{\parsep}{0pt}}{\end{itemize}

}
\usepackage{amsmath} %

\usepackage[toc,page]{appendix} %
\usepackage{wrapfig} %
\usepackage{comment}
\usepackage{cuted} %
\usepackage{lipsum} %
\usepackage{mathtools} %
\usepackage{algorithmicx}
\usepackage{algpseudocode}

\usepackage{booktabs}
\usepackage{multirow} %
\usepackage{multicol}
\usepackage{rotating} %

\usepackage{gensymb}
\usepackage{float}
\usepackage{soul}
\usepackage{array}
\usepackage{multirow}
\usepackage{hhline}
\usepackage{setspace}

\algdef{SE}[DOWHILE]{Do}{doWhile}{\algorithmicdo}[1]{\algorithmicwhile\ #1}%

\makeatletter
\renewcommand{\ALG@beginalgorithmic}{\small}
\makeatother

\newcommand{\DELETE}[1]{} %
\newcommand{\IGNORE}[1]{}
\usepackage{datenumber}
\usepackage{calc}
\usepackage[mmddyyyy]{datetime}

\newcounter{datetoday}
\newcounter{diffyears}
\newcounter{diffmonths}
\newcounter{diffdays}

\newcommand{\difftoday}[3]{%
      \setmydatenumber{datetoday}{\the\year}{\the\month}{\the\day}%
      \setmydatenumber{diffdays}{#1}{#2}{#3}%
      \addtocounter{diffdays}{-\thedatetoday}%
      \ifnum\value{diffdays}>0
        \def\diffbefore{}%
        \def\diffafter{left}%
      \else
        \def\diffbefore{}%
        \def\diffafter{ago}%
        \setcounter{diffdays}{-\value{diffdays}}%
      \fi
      \setcounter{diffyears}{\value{diffdays}/365}%
      \setcounter{diffdays}{\value{diffdays}-365*\value{diffyears}}%
      \setcounter{diffmonths}{\value{diffdays}/30}%
      \setcounter{diffdays}{\value{diffdays}-30*\value{diffmonths}}%
      \diffbefore
      \ifnum\value{diffyears}=0
      \else
        \ifnum\value{diffyears}>1
            \thediffyears\space years,
        \else
            \thediffyears\space year,
        \fi
      \fi
      \ifnum\value{diffmonths}=0
      \else
        \ifnum\value{diffmonths}>1
            \thediffmonths\space months
        \else
            \thediffmonths\space month
        \fi
      \fi
      \ifnum\value{diffdays}=0
      \else
        \ifnum\value{diffdays}>1
            \thediffdays\space days
        \else
            \thediffdays\space day
        \fi
      \fi
      \diffafter
}

\usepackage{booktabs} %
\usepackage{iccv}
\usepackage{times}
\usepackage{epsfig}
\usepackage{graphicx}
\usepackage{amsmath}
\usepackage{amssymb}
\usepackage{upgreek}

\usepackage[pagebackref=true,breaklinks=true,letterpaper=true,colorlinks,bookmarks=false]{hyperref}

\iccvfinalcopy %

\def\iccvPaperID{7653} %

\ificcvfinal\fi

\begin{document}
	
\title{Single-shot Hyperspectral-Depth Imaging with Learned Diffractive Optics}

\author{Seung-Hwan Baek\footnotemark[1]\,~\footnotemark[2]
\and
Hayato Ikoma\footnotemark[3]
\and
Daniel S. Jeon\footnotemark[1]
\and
Yuqi Li\footnotemark[4]
\and
Wolfgang Heidrich\footnotemark[4]
\and
Gordon Wetzstein\footnotemark[3]
\and
Min H. Kim\footnotemark[1] \and \\[-2mm]
	\footnotemark[1]\,~KAIST\hspace{10mm}
	\footnotemark[2]\,~Princeton University\hspace{10mm}
	\footnotemark[3]\,~Stanford University\hspace{10mm}
	\footnotemark[4]\,~KAUST
}

\twocolumn[{%
\vspace{-7mm}
\renewcommand\twocolumn[1][]{#1}%
\maketitle
\thispagestyle{empty}
}]

\ificcvfinal\thispagestyle{empty}\fi

\begin{abstract}
Imaging depth and spectrum have been extensively studied in isolation from each other for decades. Recently, hyperspectral-depth (HS-D) imaging emerges to capture both information simultaneously by combining two different imaging systems; one for depth, the other for spectrum. While being accurate, this combinational approach induces increased form factor, cost, capture time, and alignment/registration problems. In this work, departing from the combinational principle, we propose a compact single-shot monocular HS-D imaging method. Our method uses a diffractive optical element (DOE), the point spread function of which changes with respect to both depth and spectrum. This enables us to reconstruct spectrum and depth from a single captured image. To this end, we develop a differentiable simulator and a neural-network-based reconstruction method that are jointly optimized via automatic differentiation. To facilitate learning the DOE, we present a first HS-D dataset by building a benchtop HS-D imager that acquires high-quality ground truth. We evaluate our method with synthetic and real experiments by building an experimental prototype and achieve state-of-the-art HS-D imaging results.
\end{abstract}

\section{Introduction}
\label{sec:introduction}
Spectral information is crucial for a plethora of applications in the fields of remote sensing,  food/agriculture, medical imaging, and defense~\cite{nasi2015using,dale2013hyperspectral,adao2017hyperspectral,lu2014medical,briottet2006military}.
In parallel, depth imaging also has been developed for decades and now serves as a critical functionality for robotics, autonomous driving, mobile photography, and augmented/mixed reality~\cite{hansard2012time,scharstein2002taxonomy,gruber2019gated2depth}.
These two imaging modalities recently started to be merged as hyperspectral-depth (HS-D) imaging that has various applications in ornithology, geology, biology, arts, and cultural heritage~\cite{kim2012spectro,zia20153d,kitahara2015simultaneous,wu2016snapshot,feng2016compressive,ozawa2017hyperspectral,rueda2019snapshot}.
Specifically, single-shot HS-D imaging has applications such as analyzing living biological samples and geometric-spectral scene understanding from a moving camera on a vehicle or a hand-held camera.

\begin{figure}[t]
  \centering
   \vspace{-2mm}%
   \includegraphics[width=\linewidth]{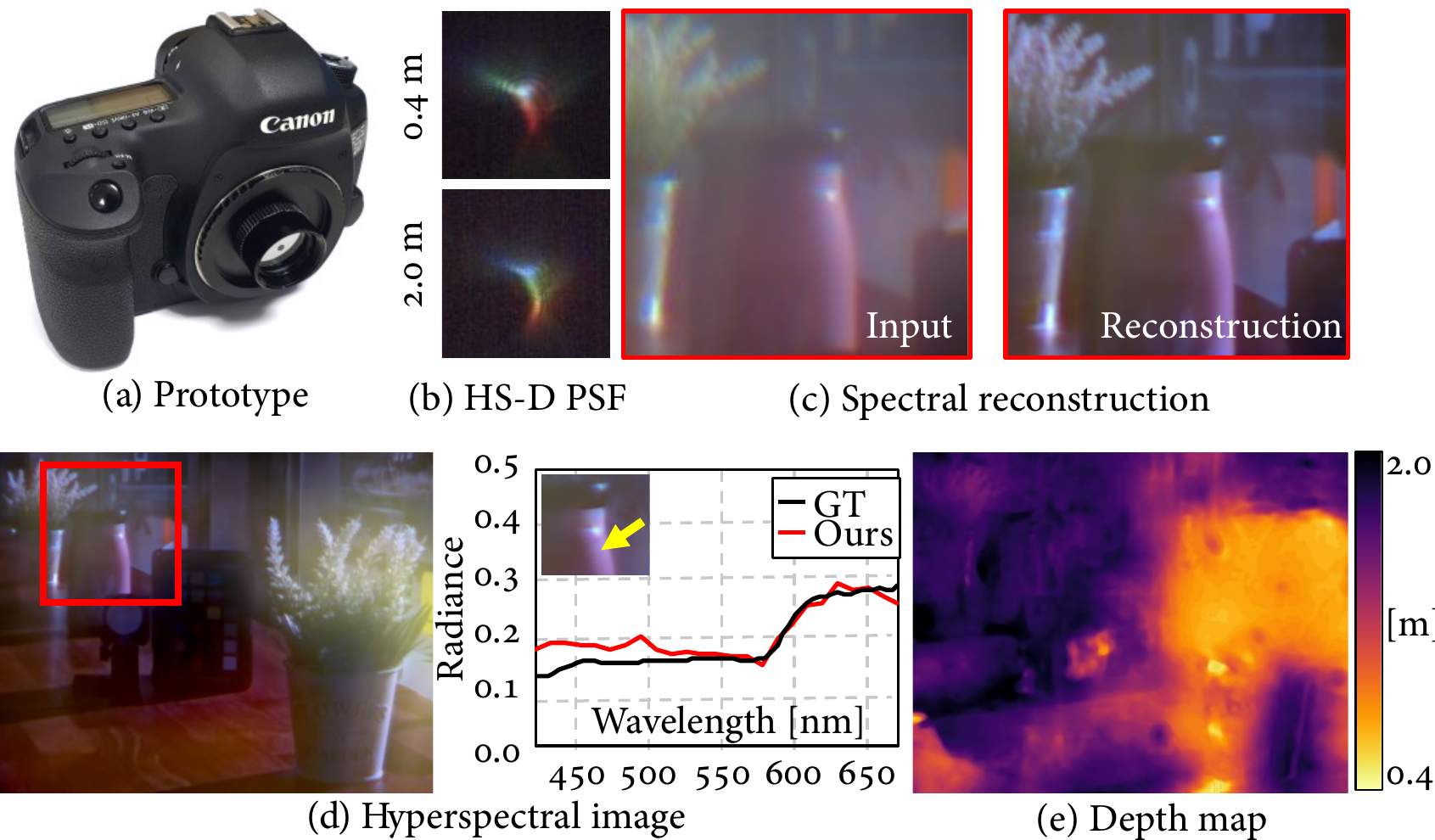}%
   \vspace{-1mm}%
   \caption[]{\label{fig:teaser}%
(a) Our compact single-shot HS-D imaging method uses an optimized DOE that creates (b) a PSF that varies with spectrum and depth.
(c)\,--\,(e) It encodes spectral-depth information in the captured image, from which we simultaneously reconstruct a depth map and a hyperspectral image from 420\,nm to 680\,nm with 10\,nm bandwidth.}
   \vspace{-2mm}%
\end{figure}

\begin{figure*}[th!]
  \vspace{-5mm}%
  \centering
  \includegraphics[width=1.0\linewidth]{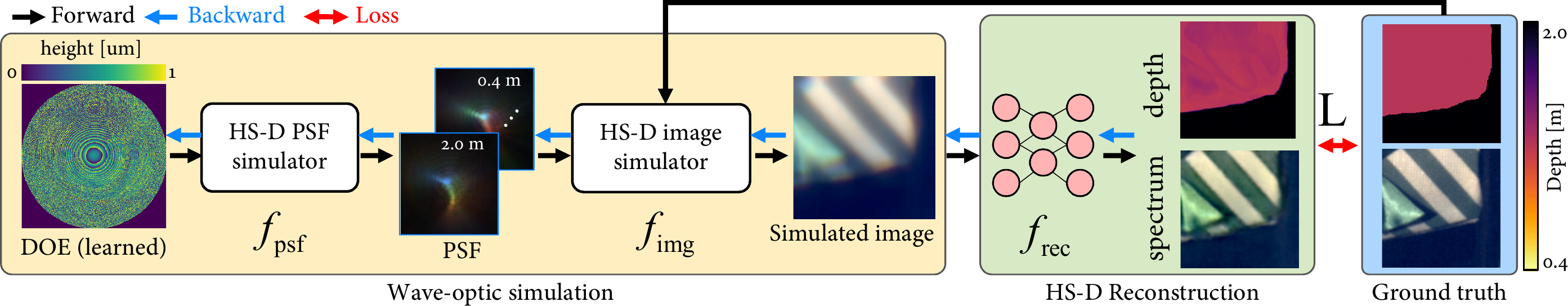}%
  \vspace{0mm}%
  \caption{\label{fig:overview}%
Our single-shot HS-D imaging is based on a differentiable pipeline that includes a wave-optics simulation and an HS-D image reconstruction part. For a DOE height map, we simulate its PSF at each depth and spectrum sample using a PSF simulator $f_\text{psf}$.
Then, the HS-D image simulator $f_\text{img}$ computes a sensor image with the aid of the ground-truth HS-D dataset.
The CNN-based reconstructor $f_\text{rec}$ estimates a hyperspectral image and a depth map.
As the entire pipeline is differentiable, we optimize the DOE and the CNN by backpropagating the loss $\mathcal{L}$.}
  \vspace{-5mm}%
\end{figure*}

To capture both spectrum and depth, existing HS-D imaging systems follow a combinational approach; they independently capture spectral and depth information with separate imaging systems, and combine the results after the fact~\cite{kim2012spectro,wang2016simultaneous,feng2016compressive,wu2016snapshot}.
This enables accurate acquisition of both data by exploiting decade-long research in each regime.
However, this combinational design fundamentally limits the potentials of HS-D imaging as it inevitably increases form factor, cost, capture time with additional alignment/registration problems.

In this work, instead of relying on two different imaging systems and combining them, we propose a first compact single-shot HS-D imaging method that uses a single diffractive optical element (DOE) in front of a conventional camera sensor.
Our key intuition is that depth and spectrum are closely coupled in DOE-based imaging systems, thus allowing for single-shot capturing of an HS-D image.
However, it is nontrivial to design a DOE that distinctively varies with scene depth and spectrum for HS-D imaging.
To solve this problem, we implement a fully differentiable image simulator that synthesizes a sensor image for a given DOE height profile, building on recent advances in automatic differentiation and Fourier optics~\cite{sitzmann2018end,Chang:2019:DeepOptics3D,wu2019phasecam3d,sun2020endtoendspad,Sun2021DiffLens,Tseng2021DeepCompoundOptics,Gordon:Nature:2020}.
Combined with a convolutional neural network (CNN) that estimates depth and spectrum from a sensor image, we build an end-to-end differentiable pipeline from the DOE profile to the reconstructed depth and spectrum, allowing us to jointly optimize DOE and CNN via backpropagation.
The obtained DOE exhibits distinct variations in its shape with respect to both depth and spectrum.

For such joint optimization, one key missing part is the ground-truth HS-D dataset to supervise the optimization.
As no such HS-D dataset exists, we build a benchtop HS-D imaging system that includes a structured light-based 3D scanning module and a bandpass filter-based hyperspectral imaging module.
With this benchtop setup, we capture a first HS-D dataset that can be used for data-driven plenoptic imaging researches.
We will make the dataset publicly available.

Our HS-D imaging method trained on the HS-D dataset outperforms the state-of-the-art single-shot HS-D imaging methods and alternative optical designs both in terms of form factor and accuracy.
Our specific contributions are as follows.
\begin{tightitemize}
\item First compact monocular HS-D imaging method with a learned DOE that captures a depth map and a hyperspectral image from a single shot,
\item First HS-D dataset of hyperspectral reflectance images and depth maps acquired by a benchtop HS-D imaging system that could fuel data-driven plenoptic imaging research, and
\item Experimental verification in simulation and real experiments by fabricating an optimized DOE.
    \end{tightitemize}

\section{Related Work}
\label{sec:relatedwork}

\vspace{0.5em}\noindent\textbf{Hyperspectral Imaging.}\hspace{0.1em}
Hyperspectral imaging has been extensively studied in the last decade.
Scanning-based approaches capture multiple 1D spectral signals by isolating the spectral energy of each wavelength from others using a set of bandpass filters, a liquid crystal tunable filter (LCTF), or a slit with dispersive optics~\cite{Brady:Book:2009}.
Compressive imaging techniques, a.k.a. coded aperture snapshot spectral imagers (CASSI), enable single-shot capture of hyperspectral images~\cite{wagadarikar2008single,jeon2016multisampling,johnson2007snapshot,Wang_2019_CVPR,DeepCASSI:SIGA:2017}.
Recent approaches have demonstrated the potential of estimating hyperspectral images from spectrally varying point spread functions (PSFs)~\cite{baek2017compact,jeon2019compact} in a compact configuration that make use of edge information instead of using the modulated aperture mask.
Our approach extends the capabilities of these spectrum-from-PSF methods by taking a first step towards snapshot imaging of higher-dimensional visual data: spectrum as well as depth.

\vspace{0.5em}\noindent\textbf{Depth Imaging.}\hspace{0.1em}
Depth imaging is a widely studied topic. The approaches closest to ours include methods using the PSFs for depth estimation from a single image \cite{BaekGutierrezKim:SIGA:2016}.
While traditional depth-from-defocus (DfD) analyzes the depth-dependent PSFs of a conventional camera to infer a depth map from two or more images \cite{nayar1994shape,subbarao1994depth},
depth-dependent spectral PSFs implemented by diffractive optical elements have also been exploited for all-in-focus particle imaging velocimetry \cite{xiong2017rainbow}.
Several groups have proposed computational photography approaches that employ amplitude-coded apertures \cite{levin2007image,veeraraghavan2007dappled} and phase masks~\cite{pavani2009three,wu2019phasecam3d} to simplify the depth estimation problem.
Our work is inspired by these approaches, but we explore applications to HS-D imaging.

\vspace{0.5em}\noindent\textbf{Hyperspectral-depth Imaging.}\hspace{0.1em}
HS-D imaging has been explored based on combinational paradigm that combines different imaging systems for spectrum and depth.
For example, passive stereo \cite{wu2016snapshot,ito20163d,zia20153d} and active stereo \cite{kitahara2015simultaneous,ozawa2017hyperspectral,kim2012spectro} have been employed in conjunction with spectral cameras \cite{wu2016snapshot,kim2012spectro,zia20153d,ozawa2017hyperspectral} and spectral light sources \cite{ito20163d,kitahara2015simultaneous}.
These approaches use two different imaging modalities for spectral and depth information, significantly increasing the device form factors and, in many cases, making it difficult to match stereo features across different spectral bands.
CASSI systems have also been combined with light-field or time-of-flight (TOF) imaging to achieve snapshot monocular imaging~\cite{feng2016compressive,rueda2019snapshot}, but these systems use custom optical coding strategies which are restricted to indoor scenes only.
To date, these systems have only been demonstrated on an optical table with a large form factor, limiting its applications.
Furthermore, parallax and related alignment problems across modalities can negatively affect the reconstruction results.
In contrast, we demonstrate a compact monocular HS-D imaging system with a learned DOE that support single-shot capability, operating in a fully passive way.

\vspace{0.5em}\noindent\textbf{Differentiable Optical Simulation.}\hspace{0.1em}
The idea of jointly optimizing optical elements with differentiable reconstruction algorithms has recently  been explored for various applications \cite{Gordon:Nature:2020}, for instance, color filter design \cite{chakrabarti2016learning}, spectral imaging  \cite{lizhi8552450}, superresolution localization microscopy \cite{nehme20deepstorm3d}, super-resolution SPAD imaging \cite{sun2020endtoendspad}, depth estimation \cite{haim2018depth,wu2019phasecam3d,Chang:2019:DeepOptics3D}, extended depth of field and super-resolution imaging \cite{sitzmann2018end}, HDR imaging~\cite{Metzler:2020:DeepOpticsHDR,Sun2020LearningRank1HDR}, and image classification \cite{chang2018hybrid}.
Based on this paradigm, we train a DOE for HS-D imaging while learning a reconstruction network.
Our approach is the first to propose and demonstrate end-to-end optimization of a single DOE and a CNN for snapshot hyperspectral and also hyperspectral-depth imaging.

\section{Diffraction-based HS-D Encoding}
\label{sec:image_formation}
Our key intuition is that the wavefront of the diffracted light wave by a DOE changes with spectrum and scene depth.
We examine this by establishing an HS-D image formation model based on Fourier optics~\cite{goodman2005introduction} in a differentiable manner.
This amounts to simulating a PSF and a sensor image for a given DOE, which comprise our pipeline shown in Figure~\ref{fig:overview}.

\begin{figure}[t]
  \centering
   \vspace{-2mm}%
  \includegraphics[width=0.9\linewidth]{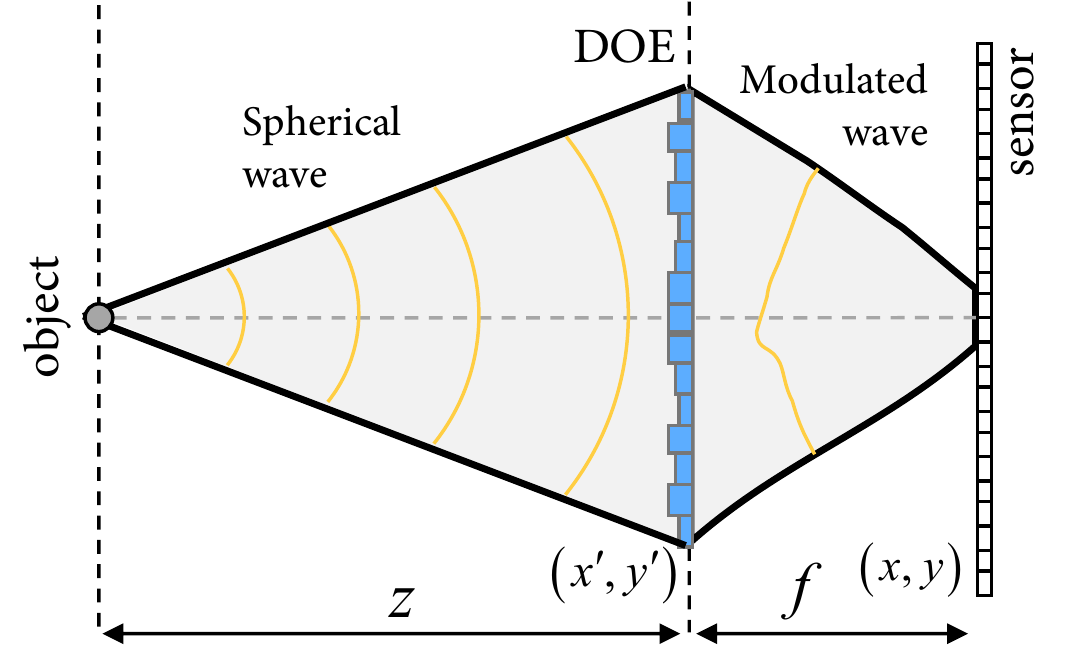}%
  \vspace{-1mm}%
  \caption[]{\label{fig:lightprop}%
   Schematic diagram of the light propagation from the object at depth $z$ to the sensor with the focal length $f$.
   The phase of the spherical light wave coming from a scene point is modulated by the DOE and captured by a conventional sensor. The corresponding PSF varies as wavelength $\lambda$ and depth $z$, enabling HS-D imaging from a single image.}
  \vspace{-4mm}%
\end{figure}

\vspace{0.5em}\noindent\textbf{Differentiable Point Spread Function.}\hspace{0.1em}
\label{sec:psf}
We denote $f_{\text{psf}}(\cdot)$ as our differentiable PSF simulator that computes a PSF for the given DOE height map $h$, wavelength $\lambda$, and depth $z$
\begin{equation}\label{eq:psf_simulator}
  P_{\lambda,z}=f_{\text{psf}}\left(h\right),
\end{equation}
where $P_{\lambda,z}$ is the PSF.
The wave field of wavelength $\lambda$ originating from the scene point at depth $z$ can be modeled as a spherical wave $U_{\lambda,z}$ in the aperture plane.
Refer to Figure~\ref{fig:lightprop}.
Under the Fresnel approximation\footnote{It assumes that the wavelength $\lambda$ is significantly smaller than the travel distance $z$: $\lambda \ll z$.}, this is modeled as
$U_{\lambda,z}^{(1)} = \mathrm{exp}\left[{i\frac{2\pi}{\lambda}\frac{{x'}^2+{y'}^2}{z}}\right]$,
where $(x',y')$ is the spatial coordinate on the DOE plane.
The wave field then passes through the camera aperture and the DOE resulting in changes of the amplitude and phase:
$U_{\lambda,z}^{(2)} = A(x',y')\cdot\mathrm{exp}\left[{i\frac{2\pi}{\lambda}\left(\frac{{x'}^2+{y'}^2}{z}+(\eta_{\lambda}-1)h(x',y')\right)}\right]$,
where $A$ is the amplitude aperture function, which is 0 for the blocked region and 1 elsewhere,
and $\eta_\lambda$ is the refractive index of the DOE material for wavelength $\lambda$.

The wave field propagates to the sensor by the focal length $f$ , resulting in the point spread function $P_{\lambda,z}$, which is the squared magnitude of the complex wave field at the sensor plane:
\begin{align}\label{eq:psf}
\begin{split}
    &P_{\lambda ,z}=| \mathcal{F}\{A\cdot\mathrm{exp}[ik(\phi_\text{scene} + \phi_\text{DOE} +\phi_\text{focal} ]\} |^2,
 \end{split}
\end{align}
where $\mathcal{F}$ is the Fourier transform, $k$ is the wave number ${{2\pi}}/{\lambda }$, and $\phi_\text{\{scene/DOE/focal\}}$
are the phase delays induced by propagating the scene to the DOE~$\phi_\text{scene}=({x'}^2+{y'}^2)/z$, the DOE itself $\phi_\text{DOE}=({\eta _\lambda } - 1)h(x',y')$, and propagating from the DOE to the sensor~$\phi_\text{focal}=({{x'}^2} + {{y'}^2})/(2f)$.

\vspace{0.5em}\noindent\textbf{HS-D Encoding in PSF.}\hspace{0.1em}
\label{sec:spectral-depth-dependency}
Equation~\eqref{eq:psf} shows that the PSF generated by a DOE depends on the wavelength~$\lambda$ and the depth $z$ of a scene point
as shown in the three phase terms $\phi_\text{scene/DOE/focal}$ augmented by the wave number $k$.
The first term $k\phi_\text{scene}$ is inversely proportional to both wavelength $\lambda$ and depth $z$,
the second term $k\phi_\text{DOE}$ is proportional to the refractive index $\eta _\lambda$ of the DOE material and inversely proportional to $\lambda$,
and the last term $k\phi_\text{focal}$ is also inversely proportional to $\lambda$ and the focal length $f$ of the DOE.
These terms are then summed up and converted to the frequency domain through Fourier transform $\mathcal{F}$.
This analysis concludes that the point spread function $P_{\lambda,z}$ changes by wavelength $\lambda$ and depth $z$.
Therefore, we can in principle reconstruct these two pieces of information from PSF analysis in a single image.

However, factorizing depth and spectrum from a single image is a severely ill-posed problem, so making traditional PSF engineering approaches rarely attempt to solve this problem.
To mitigate this challenge, we introduce a unified data-driven imaging solution that includes a learned DOE and a reconstruction network in an end-to-end manner.

\IGNORE{
\begin{figure}[t]
  \centering
   \vspace{-4mm}%
  \includegraphics[width=1.0\linewidth]{figs/psf.pdf}%
  \vspace{-1mm}%
  \caption{\label{fig:psf}%
An example PSF of a Fresnel DOE that changes as depth and spectrum.
The isotropic shape of the PSF and spectral-depth ambiguity along a diagonal direction make the Fresnel DOE suboptimal for HS-D imaging.}
\vspace{-2mm}
\end{figure}
}

\begin{figure*} [t]
  \centering
\includegraphics[width=1.0\linewidth]{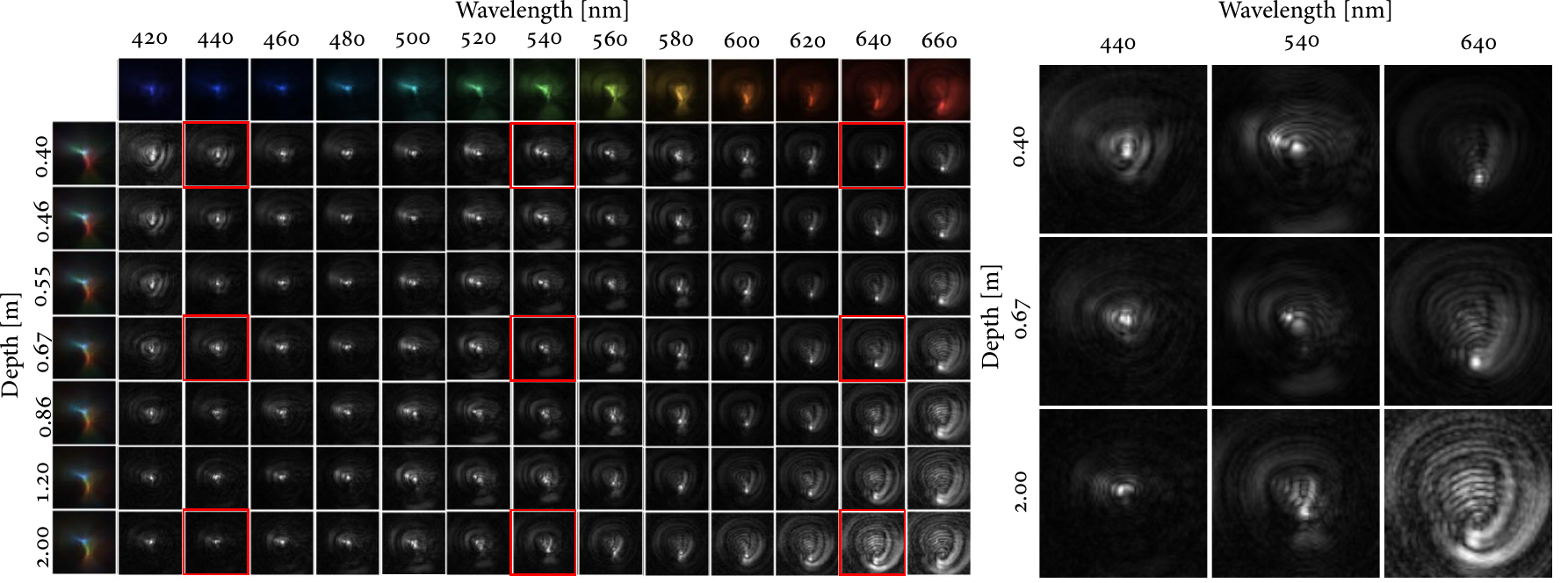}%
  \vspace{-0mm}%
  \caption{\label{fig:init}%
  Our optimized PSF changes with respect to spectrum and depth over the spectral range from 420 to 660\,nm and the depth range from 0.4 to 2.0\,m.
  It enables the single-shot HS-D capture using a  reconstruction network, simultaneously trained in the end-to-end manner.}
  \vspace{-4mm}
\end{figure*}

\begin{figure}[t]
  \centering%
  \includegraphics[width=1.0\linewidth]{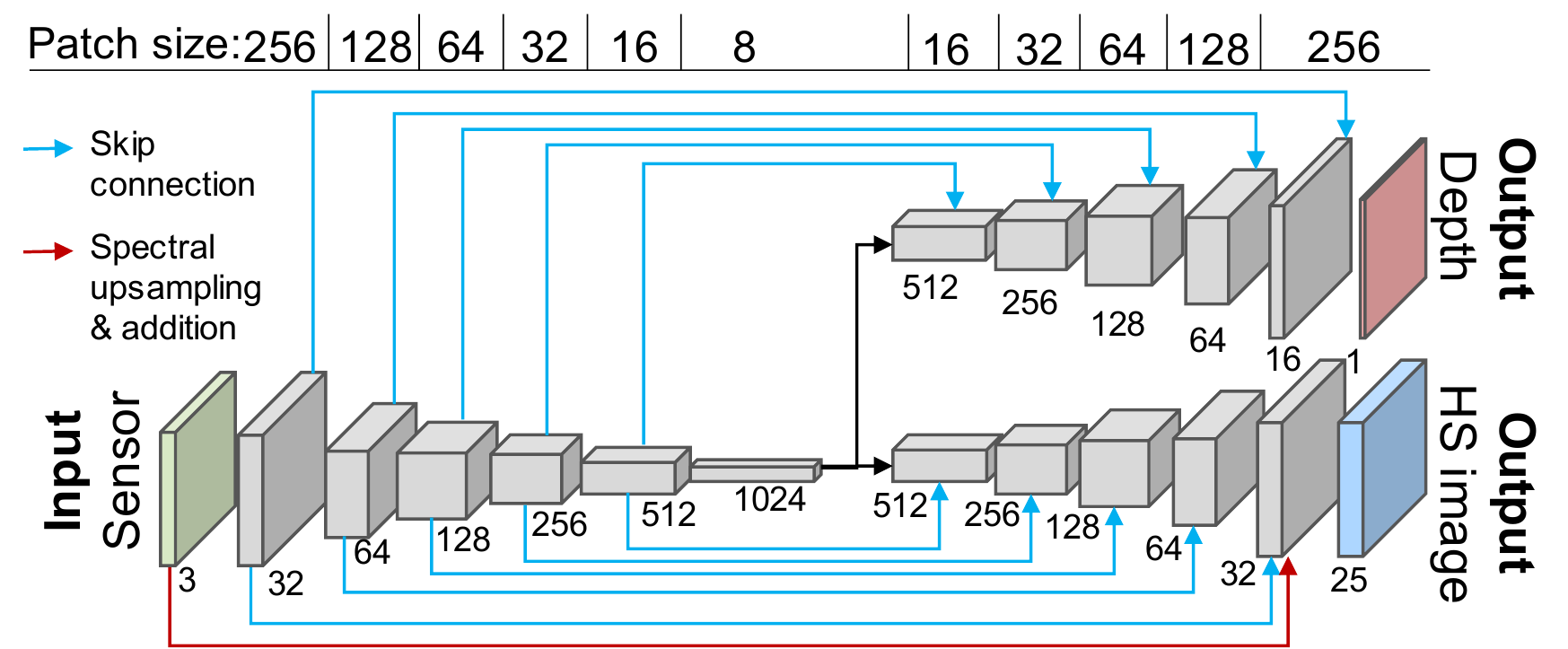}%
  \vspace{-0mm}%
  \caption{\label{fig:network}%
  We take the sensor RGB image and extracts features using an encoder. Two  decoders convert the features to a hyperspectral image and a depth map, respectively. In addition to the skip connection between the encoder and the decoders, we implement residual learning via a spectral-upsampling module.
  }%
  \vspace{-4mm}
\end{figure}

\begin{figure}[t]
  \centering
  \includegraphics[width=1.0\linewidth]{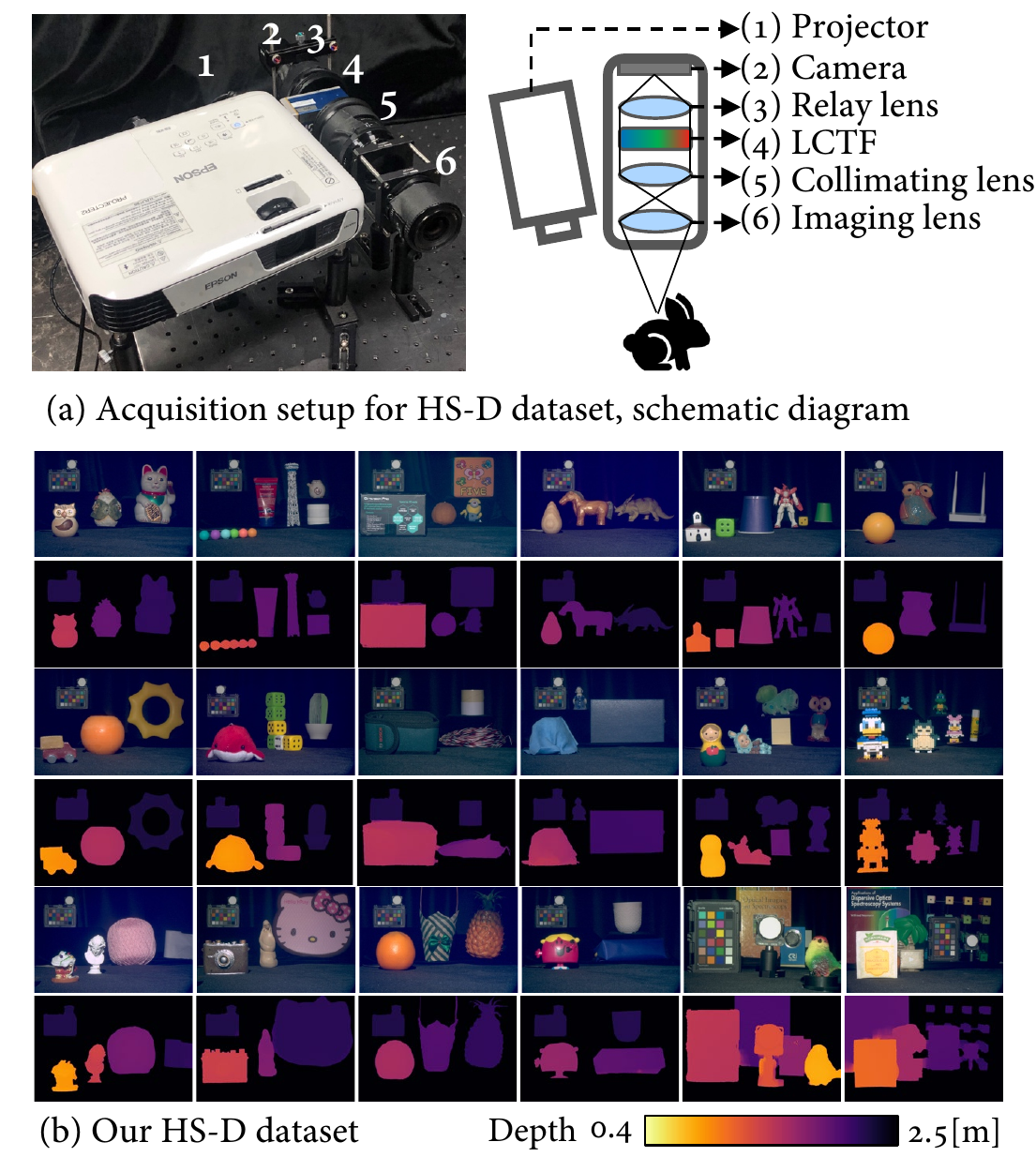}%
  \vspace{-2mm}%
  \caption{\label{fig:dataset}%
  (a) We built a benchtop HS-D imaging setup using structured-light 3D scanning and LCTF-based hyperspectral imaging.
  (b) We capture a first HS-D dataset consisting of hyperspectral reflectance images and depth maps using the benchtop setup.}
  \vspace{-4mm}%
\end{figure}

\IGNORE{%
Figure~\ref{fig:psf} shows the simulated PSFs of a Fresnel DOE with varying wavelength and depth.
We observe the predicted spectrum--depth ambiguity along the diagonal lines,
e.g., the PSF of low wavelength at a far distance is similar to the PSF of long wavelength at a near distance.
A traditional PSF engineering approach would likely aim at directly optimizing the properties of the PSF to mitigate this ambiguity, which is not trivial and guaranteed to improve the final reconstruction quality of depth and spectrum.
Instead, we optimize the DOE in an end-to-end manner together with the reconstruction network. This approach allows us to place the loss directly on the estimated HS-D data, rather than on arbitrary proxy metrics computed for the PSF.
}%

\vspace{0.5em}\noindent\textbf{Sensor Image Synthesis.}\hspace{0.1em}
\label{sec:image-formation}
Given the PSF~$P_{\lambda,z}$, a pair of all-in-focus hyperspectral image $I_{\lambda}$ and a depth map $Z$ from our HS-D dataset, we simulate the corresponding sensor image $J_{c\in\{R,G,B\}}$ using an image simulator $f_{\text{img}}(\cdot)$.
It is based on a convolutional PSF model and a layered scene representation, where a scene is modeled as a composition of multiple layers at different depths~\cite{Chang:2019:DeepOptics3D,wu2019phasecam3d}.
The sensor image $J_{c}$ is then modeled as follows:
\begin{align}\label{eq:image_formation}
J_{c}&=f_{\text{img}}\left(P_{\lambda,z},I_{\lambda},Z\right)  \nonumber \\
&= \sum\limits_{\lambda  \in \Lambda } \Omega_{c,\lambda}{\sum\limits_{z \in Z} {{M_z} \odot \left( {{I_\lambda } \otimes {P_{\lambda ,z}}} \right)} } + n,
\end{align}
where $\Omega_{c\in\{R,G,B\},{\lambda\in\Lambda}}$ is the camera response function, $\odot$ is an element-wise product operator, $\otimes$ is a convolution operator, and $M_z$ is the weight map for each depth layer $z$.
We compute $M_z$ by applying a Gaussian filter to the binarized occupancy map that has value of one if the pixel depth is at~$z$, and zero otherwise~\cite{Chang:2019:DeepOptics3D,wu2019phasecam3d}.
Lastly, we apply signal-independent additive Gaussian noise $n$.%
We exclude Poisson noise due to its non-differentiability and even its reparameterization introduces training instability~\cite{paszke2017automatic}.

\vspace{0.5em}\noindent\textbf{HS-D Sampling.}\hspace{0.1em}
\label{sec:sampling}
We use dense sampling of 25 spectral channels from 420 to 660\,nm in 10\,nm intervals and seven depth levels from 0.4 to 2.0\,m linearly spaced in disparity.
This is essential not only for estimating an HS-D image from the spectral cue of PSFs, but also for accurately simulating the image formation model.

\section{End-to-End HS-D Reconstruction}
\label{sec:e2e-optimization}

The spectrum-depth dependency of the PSF allows us to reconstruct a spectral image $\hat{I}_\lambda$ and a depth map $\hat{Z}$ from a captured image $J_{c}$. We use a deep convolutional neural network for the reconstruction algorithm $f_{\text{rec}}(\cdot)$:
\begin{equation}\label{eq:reconstructor}
  \hat{I}_\lambda,\hat{Z} = f_{\text{rec}}\left(J_{c}\right).
\end{equation}

\vspace{0.5em}\noindent\textbf{Network Architecture.}\hspace{0.1em}
\label{sec:network}
We design our network architecture based on the U-Net architecture~\cite{ronneberger2015u} with  a dual-decoder design: one for depth and the other for the HS image (Figure~\ref{fig:network}). %
This design is memory efficient essential for our end-to-end learning that consumes vast amount of GPU memory for differentiable HS-D simulation.
Further network details can be found in the Supplemental Document.

\vspace{0.5em}\noindent\textbf{Residual Learning with Spectral Upsampling.}\hspace{0.1em}
To reduce the ill-posedness of the reconstruction problem,
we apply residual learning~\cite{Gharbi:2016:DJD:2980179.2982399} by adding the initial spectral tensor to the output HS image.
However, different from conventional residual learning for super-resolution and demosaicing, the number of output spectral channels is larger than the input channels: from 3 to 25.
As such, we propose to distribute the input RGB-channel intensity to the hyperspectral channels based on the camera response function $\Omega$:
${I_\lambda^{\text{up}} } = \sum\nolimits_{c} {w\left( {\lambda ,c} \right){J_c}}$, where $w\left( {\lambda ,c} \right) = {{\Omega (\lambda ,c)} \mathord{\left/
 {\vphantom {{\Omega (\lambda ,c)} {\sum {\Omega (\lambda , \cdot )} \sum {\Omega ( \cdot ,c)} }}} \right.
 \kern-\nulldelimiterspace} \{ \sum\nolimits_{c'} {\Omega (\lambda , c' )} \sum\nolimits_{\lambda'} {\Omega (\lambda' , c )}  \}}$ and $c,c' \in \{r,g,b\}$.
The upsampled image $I_\lambda^{\text{up}}$ is added to the output of the hyperspectral decoder, enabling effective residual learning (the red arrow in Figure~\ref{fig:network}).

\vspace{0.5em}\noindent\textbf{Loss Function.}\hspace{0.1em}
\label{sec:loss}
Our loss function $\mathcal{L}$ is defined as the mean absolute error (MAE) of the inverse depth $D=1/Z$ and the hyperspectral image $I_{\lambda}$, in addition to the total variation regularizer on depth:
\begin{align}\label{eq:loss}
  {\mathcal L} = \alpha\frac{1}{N}{\left\| {{{\hat I}_\lambda } - {I_\lambda }} \right\|_1} + \beta\frac{1}{M}{\left\| {\hat D - D} \right\|_1} + \gamma \frac{1}{M}\left\| \nabla D\right\|_1,
\end{align}
where $N$ and $M$ are the number of total pixels in the hyperspectral image and the depth map, respectively, and $\nabla$ is the spatial gradient operator.
$\alpha=1$, $\beta=10^{-2}$, and $\gamma=10^{-2}$ are the balancing weights.

\begin{figure*}[t]
  \centering
  \vspace{-5mm}%
  \includegraphics[width=1\linewidth]{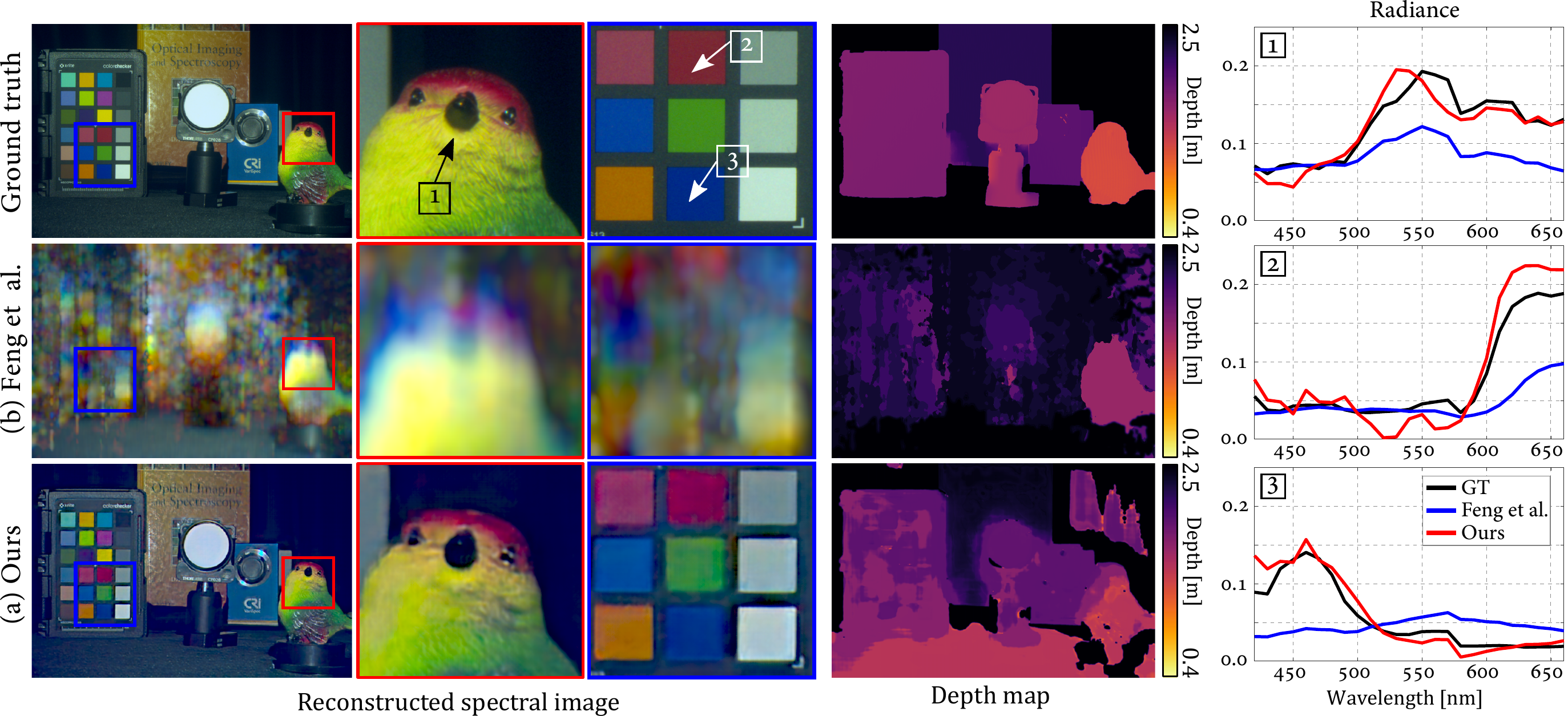}%
  \vspace{-0mm}%
  \caption{\label{fig:hsd_comp}%
  The spatial resolution of the spectral image and the depth map reconstructed by Feng et al.~\cite{feng2016compressive} is very low due to the fundamental architecture of light-field-based HS-D imaging.
  In contrast, our HS-D imaging can capture the spectral image and depth information with higher accuracy
  while our DOE-based camera has a significantly smaller form factor than the state-of-the-art system.}
  \centering
  \vspace{4mm}%
  \includegraphics[width=0.99\linewidth]{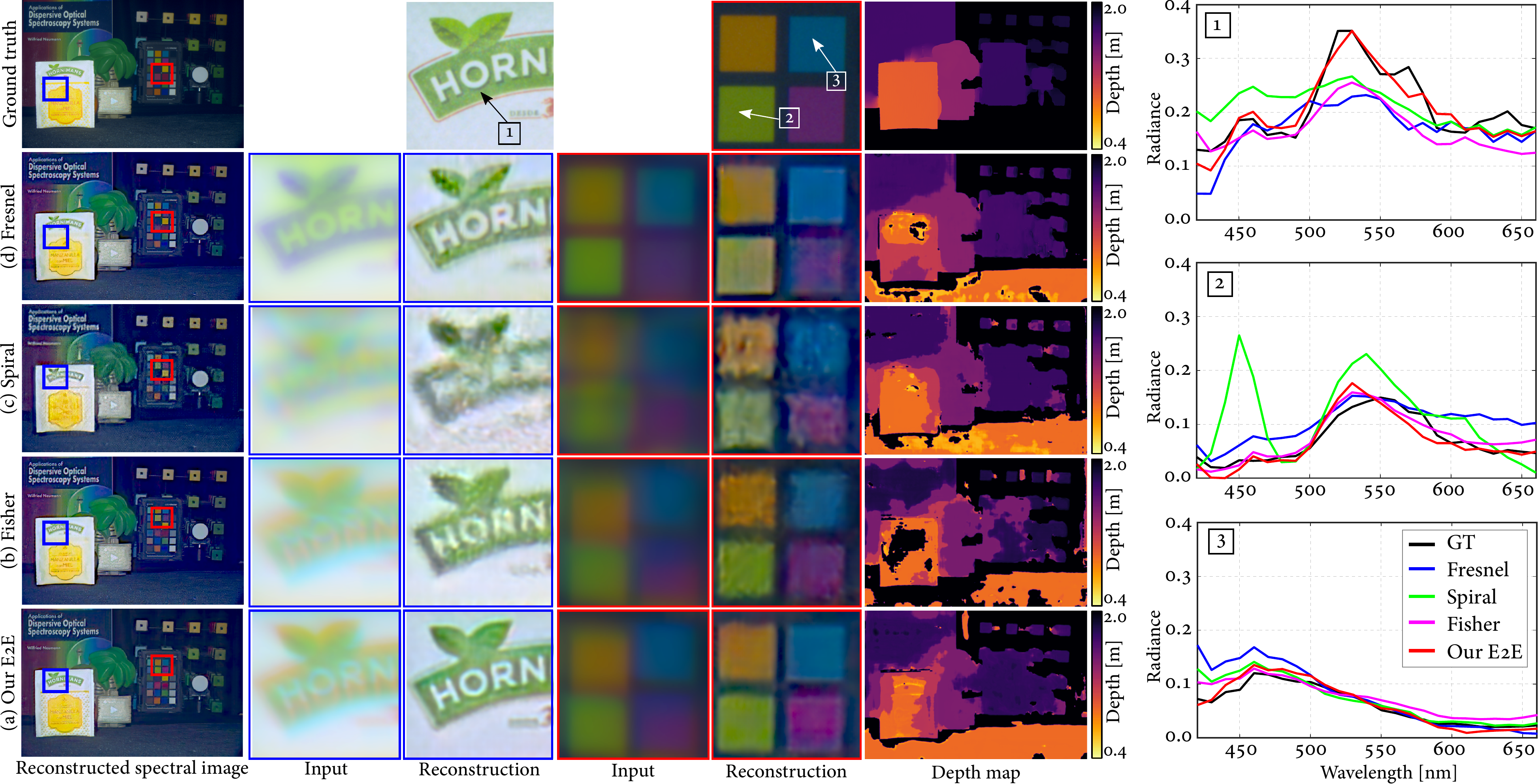}%
  \vspace{-0mm}%
  \caption{\label{fig:e2e_comparison}%
  We compare our learned DOE with the Fresnel, the Spiral~\cite{jeon2019compact}, the Fisher~\cite{shechtman2014optimal} DOEs.
  The spectral image and depth map reconstructed by our  DOE method present significant improvements against results of the analytically-driven DOEs.}
  \vspace{-5mm}
\end{figure*}

\begin{table}[t]
\centering
	\resizebox{0.65\columnwidth}{!}{%
\begin{tabular}{c|c|r|r}\hline
\multicolumn{2}{c|}{HS-D imaging} & Feng et al.~\cite{feng2016compressive} & Ours \\ \hline \hline
\multirow{2}{3mm}{\begin{sideways}{\small{Spec.}}\end{sideways}}
& PSNR {[}dB{]} & 23.62 & \textbf{29.31} \\
& SSIM  &  0.76 &  \textbf{0.81} \\ \hline
\multirow{2}{3mm}{\begin{sideways}{\small{Depth}}\end{sideways}}
& RMSE {[}m{]} & 0.57 &  \textbf{0.20}   \\
& MAE {[}m{]} &  0.30 &  \textbf{0.12}   \\ \hline
\end{tabular}}
\vspace{1mm}%
\caption{\label{tab:comp_hsdimaging}%
{Our method outperforms the state-of-the-art HS-D imaging method~\cite{feng2016compressive} that combines the light-field imaging and the compressive spectral sensing.}
}
\vspace{-2mm}
\end{table}

\begin{table}[t]
\resizebox{\columnwidth}{!}{%
\begin{tabular}{c|c|r|r|r|r}\hline
\multicolumn{2}{c|}{DOE} & Fresnel & Spiral~\cite{jeon2019compact} & Fisher~\cite{shechtman2014optimal} & E2E (ours) \\ \hline \hline
\multirow{2}{3mm}{\begin{sideways}{\small{Spec.}}\end{sideways}}
& PSNR {[}dB{]} & 27.96 & 26.90 & {28.51} & \textbf{29.31} \\
& SSIM  & 0.74 & 0.64 & {0.79} & \textbf{0.81} \\ \hline
\multirow{2}{3mm}{\begin{sideways}{\small{Depth}}\end{sideways}}
& RMSE {[}m{]} & {0.21} & 0.32 & 0.23 & \textbf{0.20}   \\
& MAE {[}m{]} & {0.15} & 0.20 & {0.15} & \textbf{0.12}   \\ \hline
\end{tabular}}
\vspace{0mm}
\caption{\label{tab:comp_psfs}%
  Our optimized DOE outperforms the alternative DOE designs (Fresnel, Spiral~\cite{jeon2019compact}, Fisher~\cite{shechtman2014optimal}) for HS-D imaging. }
\vspace{-3mm}
\end{table}

\begin{figure*}[t]
  \centering
  \vspace{-2mm}%
  \includegraphics[width=\linewidth]{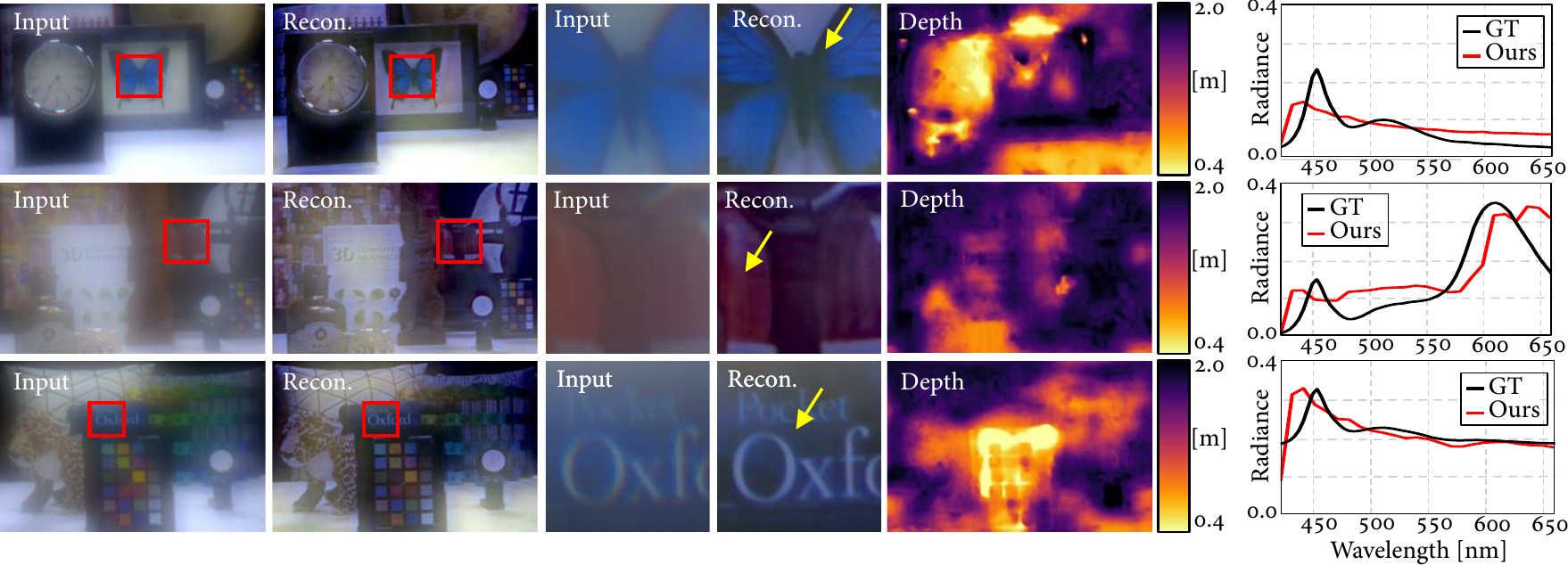}%
  \vspace{-2mm}%
  \caption{\label{fig:real_results}%
  Reconstructed hyperspectral-depth images of real-world, casual scenes. We captured these scenes with our prototype and compare the normalized radiance of resulting HS-D data with the ground-truth measured by a spectroradiometer (SpectraScan 655) at points indicated by yellow arrows.}
  \vspace{-2mm}
\end{figure*}

\begin{figure}[pt]
  \centering
  \includegraphics[width=1.0\linewidth]{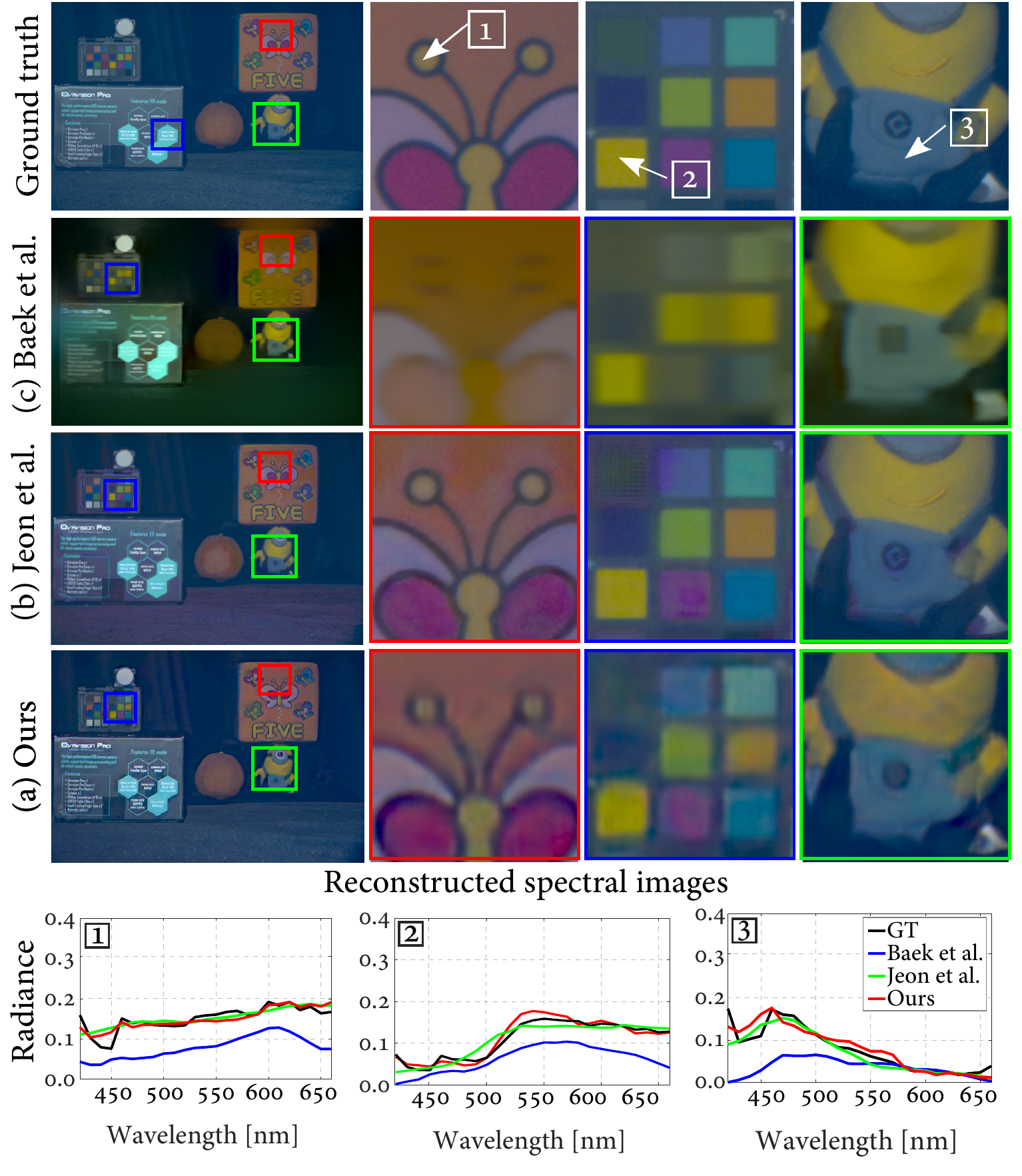}\\[0mm]
\resizebox{1.0\columnwidth}{!}{%
  \begin{tabular}{c|r|r|r}
    \hline
HS imaging   & Baek et al.~\cite{baek2017compact} & Jeon et al.~\cite{jeon2019compact} & Ours \\  \hline \hline
HS PSNR [dB] & 27.96 & 28.81  & \textbf{29.31} \\
HS SSIM & 0.75 & \textbf{0.81}  & \textbf{0.81} \\ \hline
Luminance PSNR [dB] & 28 & \textbf{40}  & 32 \\
Luminance SSIM & 0.91 & \textbf{0.96}  & 0.94 \\ \hline
\end{tabular}}\\[1mm]
  \caption{\label{fig:hs_comp}%
  We compare our method with the state-of-the-art HS imaging methods~\cite{baek2017compact,jeon2019compact}.
  Our proposed method outperforms both approaches in spectral accuracy (PSNR and SSIM computed on the hyperspectral cube) while achieving second-best performance in terms of spatial structure (PSNR and SSIM computed on the luminance image of the hyperspectral cube).
  }
  \vspace{-5mm}
\end{figure}

\vspace{0.5em}\noindent\textbf{Training.}\hspace{0.1em}
\label{sec:training}
As the HS-D simulation and the reconstruction network are both differentiable, we jointly optimize the DOE and the network by solving a minimization problem via backpropagation:
\begin{equation}\label{eq:opt}
\mathop {{\rm{minimize}}}\limits_{h ,\theta} {\mathcal L}\left( {\{ {\hat Z\left( {h ,\theta} \right),{{\hat I}_\lambda }\left( {h ,\theta} \right)}\},\left\{ {Z,{I_\lambda }} \right\}} \right),
\end{equation}
where  $h$ is the DOE height, $\theta$ is a set of network parameters.
We use patch-wise training with the patch size of 256$\times$256.
The DOE height $h$ is initialized by Fisher information, the details of which are provided in the Supplemental Document.

\vspace{0.5em}\noindent\textbf{Learned PSFs.}\hspace{0.1em}
If HS-D PSFs contain too complicated structures in spectral and depth dimensions, this inevitably leads to large spatial occupancy, limiting accurate deconvolution.
If HS-D PSFs do not provide discriminative features, we suffer from metameric ambiguity of spectrum and depth. Our learned PSFs shown in Figure~\ref{fig:init} are computationally obtained and exhibit spectral-depth variations with a compact size. Combining our neural network and the PSFs, we exploit both low-level optical PSF cue and higher-level contextual image cue to reconstruct a hyperspectral image and a depth map.

\section{HS-D Dataset}
\label{sec:training}
To supervise the training, we present a \textit{first} HS-D dataset.
We provide 18 aligned pairs of a hyperspectral reflectance $R_\lambda$ and a depth map $Z$ (see Figure~\ref{fig:dataset}).
Objects are carefully placed to avoid occlusion of the structured illumination.
The reflectance $R_\lambda$ is \textit{spectrally} augmented to radiance $I_\lambda$ using 29 CIE standard illuminants, resulting in 522 hyperspectral images.
Our dataset covers visible spectrum from 420\,nm to 680\,nm with 10\,nm interval and depth is in the range from 0.4\,m to 2.5\,m.
Clear object boundary is guaranteed through manually-obtained background masks, assuring patch sampling from valid regions.
For this data acquisition, we present a combinational HS-D imager combining structured light and LCTF-based hyperspectral imaging that acquires accurate spectral and depth data using brute-force scanning (Figure~\ref{fig:dataset}).
Patch-wise training is employed for data efficiency, and 20,000 distinctive patches are used in total.
Details on the imager can be found in the Supplemental Document.

\section{Results}
\label{sec:results}
We perform synthetic evaluation of our method compared to the state-of-the-art methods and experimentally validate our prototype.

\vspace{0.5em}\noindent\textbf{HS-D Imaging.}\hspace{0.1em}
Previous HS-D imaging methods take multi-sampling strategy in temporal domain~\cite{leitner2015hyperspectral}, with multiple sensors~\cite{xiong2017snapshot,zhu2018hyperspectral}, or direct assorted filtering~\cite{cui2020snapshot}.
One notable exception is the work~\cite{feng2016compressive} that aims single-shot HS-D imaging by combining compressive spectral imaging and light-field imaging.
Their system form factor is significantly larger than ours due to the complicated elements including a coded aperture mask, relay lenses, a prism, and a microlens array.
Our method, with a much compact size, outperforms their method~\cite{feng2016compressive} as shown in Table~\ref{tab:comp_hsdimaging} and Figure~\ref{fig:hsd_comp}.

\vspace{0.5em}\noindent\textbf{DOE Design.}\hspace{0.1em}
We compare our optimized DOE with alternative DOE designs: a Fresnel DOE, a spiral DOE~\cite{jeon2019compact}, and a Fisher DOE~\cite{shechtman2014optimal}.
For fair comparison, we use our reconstruction network for all DOE designs.
Table~\ref{tab:comp_psfs} and Figure~\ref{fig:e2e_comparison} show that our DOE design outperforms the analytically driven designs.
The Fresnel-lens DOE suffer from spectral metamerism with washed color artifacts (Figure~\ref{fig:e2e_comparison}d).
The Fisher-information DOE and spiral DOE result in degraded spatial resolution (Figures~\ref{fig:e2e_comparison}b \&~\ref{fig:e2e_comparison}c).
While quantitative difference may appear small, the impact is clearly visible in qualitative reconstruction where high-frequency structures are preserved in our results.

\begin{figure}[tp]
  \centering
  \includegraphics[width=\linewidth]{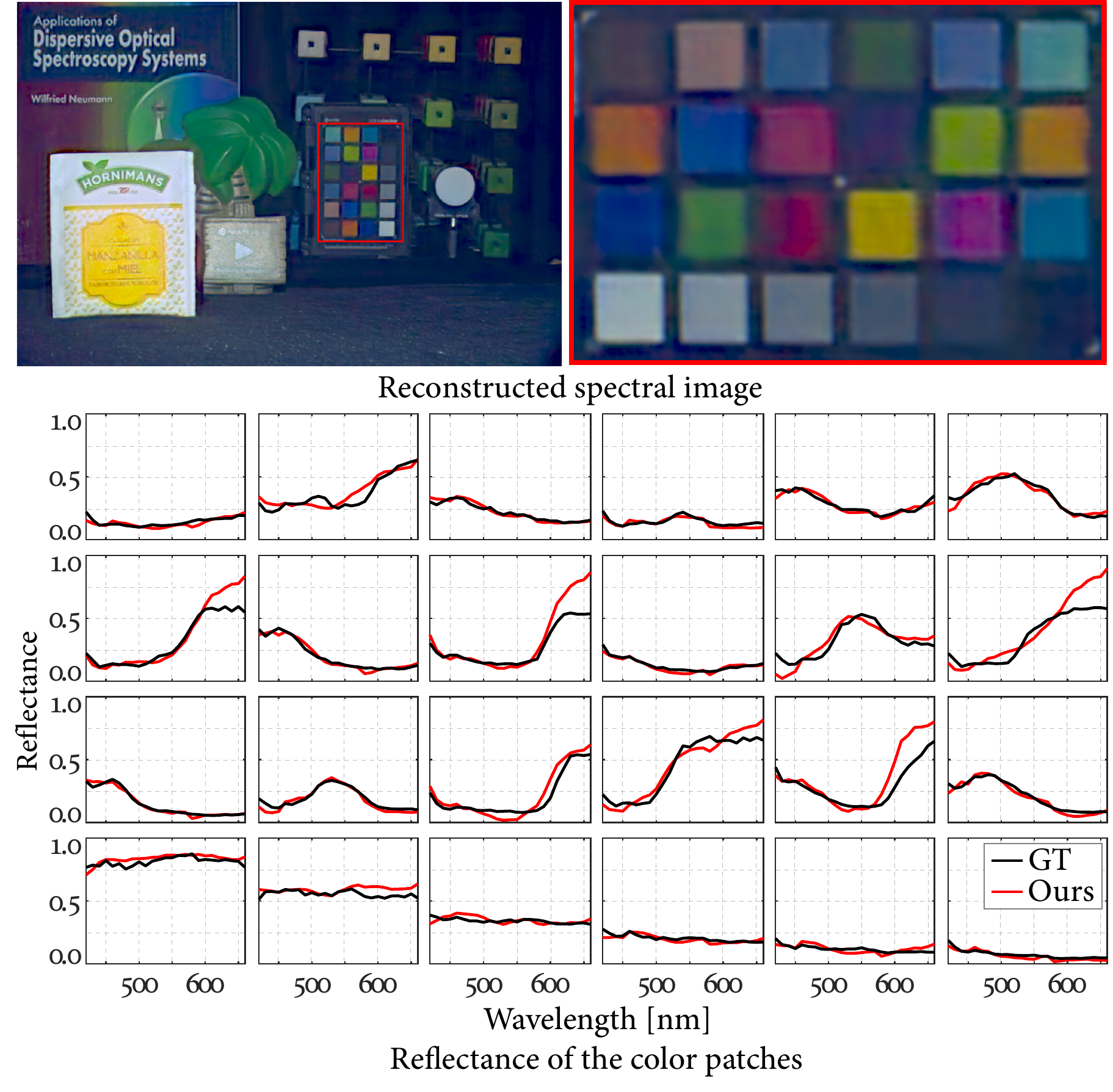}%
  \vspace{-1mm}
  \caption{\label{fig:colorchecker}%
  Quantitative evaluation of our HS-D imaging with a ColorChecker in a simulation with the ground truth. Spectral plots of 24 patches in the ColorChecker present spectral reconstruction of our method with high accuracy.}
  \vspace{-4mm}
\end{figure}

\vspace{0.5em}\noindent\textbf{Hyperspectral Imaging.}\hspace{0.1em}
Figure~\ref{fig:hs_comp} compares our system with two hyperspectral-only imaging systems: a spiral DOE-based method \cite{jeon2019compact}
and a prism-based spectral imaging method \cite{baek2017compact}.
It clearly shows that our method outperforms both methods.
Baek et al.~\cite{baek2017compact} suffers from smooth edge structures.
Jeon et al.~\cite{jeon2019compact} produces overly smoothed spectral variation.
Furthermore, different from their methods, our method estimate a hyperspectral image as well as a depth map.
Refer to the supplemental document for the experimental details.

\begin{figure}[t]
  \centering
  \includegraphics[width=\linewidth]{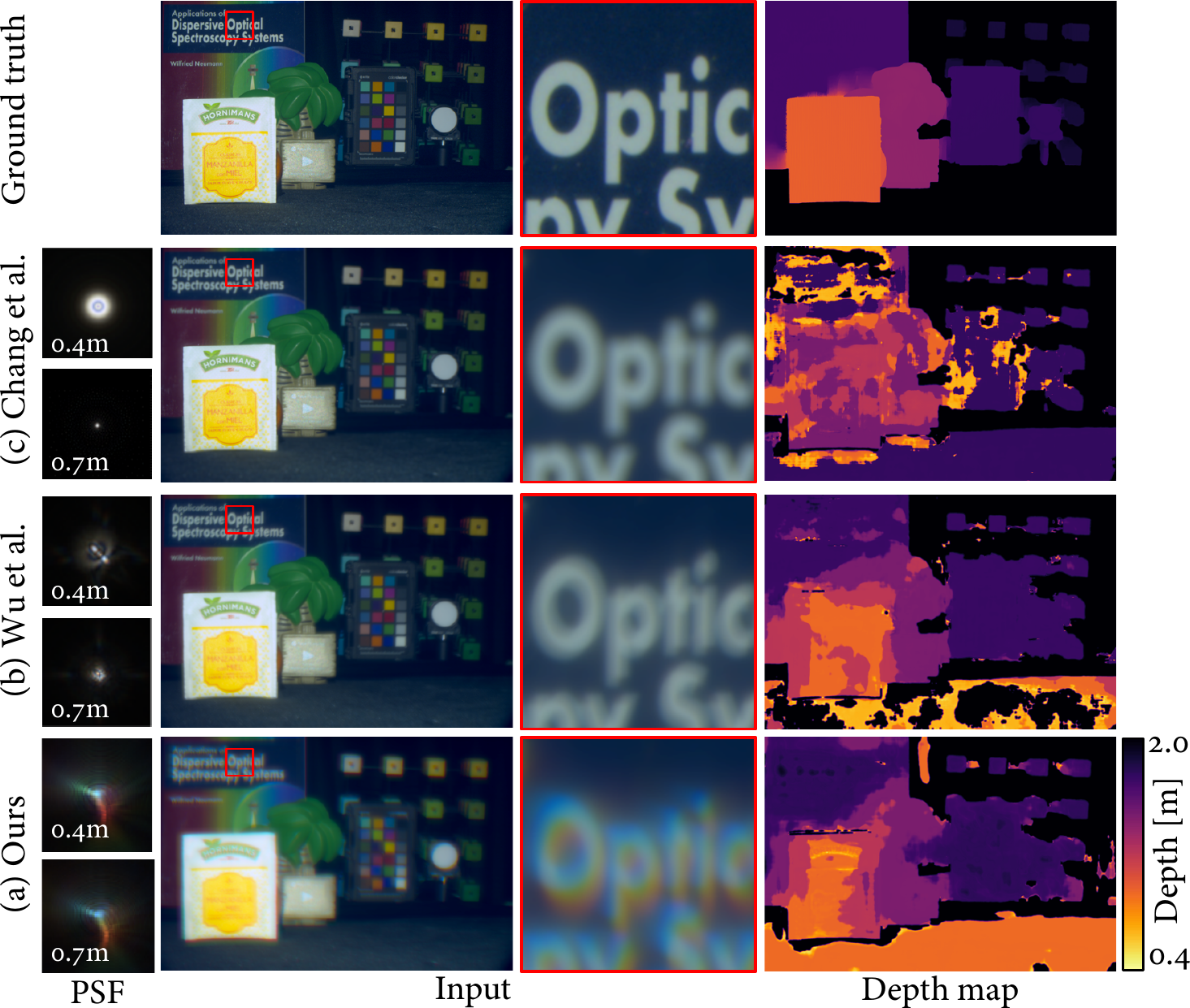}\\[1mm]
  \resizebox{0.9\columnwidth}{!}{%
    \begin{tabular}{c|c|c|c}
    \hline
      Depth imaging & Chang et al.~\cite{Chang:2019:DeepOptics3D} & Wu et al.~\cite{wu2019phasecam3d} & Ours \\
      \hline \hline
      RMSE [m] & 0.45 & 0.31 & \textbf{0.20} \\
      MAE [m] & 0.25 & 0.19 & \textbf{0.12} \\
      \hline
    \end{tabular}}\\[1mm]
  \caption{\label{fig:depth_comp}%
  We compare our method with the state-of-the-art DOE-based depth imaging methods~\cite{Chang:2019:DeepOptics3D,wu2019phasecam3d}.
Note that these methods are not designed for hyperspectral imaging.
  Our learned DOE outperforms the other two alternative designs in terms of depth accuracy.}
\vspace{-2mm}
\end{figure}

\vspace{0.5em}\noindent\textbf{Depth Imaging.}\hspace{0.1em}
We compare our method to two DOE-based depth-only imaging methods: Wu et al.~\cite{wu2019phasecam3d} and Chang et al.~\cite{Chang:2019:DeepOptics3D}, by retraining our reconstruction network for each PSF.
Figure~\ref{fig:depth_comp} shows that our method can reconstruct clearer depth maps than other methods while it also reconstruct spectral information in addition.
Refer to the supplemental document for the experimental details.

\vspace{0.5em}\noindent\textbf{PSF Dependency on Incident Angle.}\hspace{0.1em}
We found that our learned PSF exhibits insignificant variation with respect to incident angle.
For validation, we simulate two PSFs at zero and eight degrees of incident angles, where the eight degree amounts to $\sim$60$\%$ of the vertical FOV (27\,degrees).

\vspace{0.5em}\noindent\textbf{Discriminating Power of PSF.}\hspace{0.1em}
We evaluate the discriminating powers of DOE using Cramer-Rao Lower Bound (CRLB)~\cite{shechtman2014optimal}.
The CRLBs of the Fresnel/Spiral/Fisher DOEs are 2.73/1.89/1.36.
As the Fisher DOE provides the highest discriminating power, we use this as an initialization point.
However, the CRLB metric was originally designed only for a single sample at a 3D location and an impulse spectral peak.
As we aim to capture natural scenes, not a single point, our end-to-end learning optimizes the DOE with the initialization from the Fisher DOE.

\vspace{0.5em}\noindent\textbf{Spectral Evaluation.}\hspace{0.1em}
In order to evaluate the spectral accuracy of our system, we compare the reconstructed radiance of 24 patches in the standard ColorChecker under the CIE D65 illuminant with the ground truth in the simulation.
As shown in Figure~\ref{fig:colorchecker}, our results closely match the spectral power distributions of every patch in the ground truth data,
although we intentionally excluded the ColorChecker images in the training process to avoid overfitting of the network parameters to this target.
The mean RMSE of reflectance (0.0--1.0) of all 24 patches is just 0.0478.

\vspace{0.5em}\noindent\textbf{HS-D Imaging Prototype.}\hspace{0.1em}
\label{sec:psf-calibration}
We build our HS-D camera prototype using a Canon 5D Mark III camera sensor that has a pixel pitch of 6.22\,$\mu$m and a resolution of 3840$\times$5760 pixels.
The focal length of the DOE is 50\,mm.
We fabricate the optimized DOE through soft lithography~\cite{xia1998soft} of which detail is in the Supplemental Document.
The DOE is then mounted to a C-mount tube, which is attached to the camera body with an EOS-C adapter.
We made an additional C-mount extender with a 3D printer to place the DOE at the exact distance from the sensor.
Calibration details, finetuning with the real PSFs, and the mismatch between the real-simulation PSFs are in the Supplemental Document.
With our compact experimental prototype, we captured five real-world scenes shown in Figures~\ref{fig:teaser} and~\ref{fig:real_results}, captured under sunlight and office-led lighting respectively.
Ground-truth spectrum and depth for scene points are measured using a spectroradiometer and a laser distance meter.

\vspace{0.5em}\noindent\textbf{Limitations.}\hspace{0.1em}
Our single-shot HS-D imaging method struggles with low-light and intensity-saturated scenes.
Further, long-tail PSFs of the fabricated DOEs result in low contrast for real captured images, which is a prevailing artifact of existing DOE-based imaging methods due to the fabrication inaccuracy.
A network architecture with more channels and depths would be beneficial to improve reconstruction quality.
However, this is prohibited in our method due to the HS-D image simulation as a memory hog.
Our method customizes DOEs for a target camera configuration. Its generalization to various camera parameters leaves as an interesting future work.
Lastly, incorporating reconfigurable multiple DOEs could enhance depth and spectral accuracy.

\section{Conclusion}
\label{sec:conclusion}
We have presented a single-shot HS-D imaging system that consists of a learned DOE and a conventional DSLR camera.
In contrast to the existing combinational approaches,
the spectral and depth dependency of PSF is carefully analyzed and exploited in an end-to-end learning manner.
To enable the joint learning procedure, we created the first HS-D dataset.
Various experiments with the simulation and the real system validate the quality and accuracy of our method.
Also, the proposed method can be used to reduce the form factor of HS-D imaging significantly, enabling compact and portable HS-D imaging without compromising quality and accuracy.

\section*{Acknowledgements}
\label{sec:acknowledgements}
Min H.~Kim acknowledges the support of Korea NRF grant (2019R1A2C3007229) in addition to  Samsung Electronics, MSIT/IITP of Korea (2017-0-00072), National Research Institute of Cultural Heritage of Korea (2021A02P02-001), and Samsung Research Funding Center (SRFC-IT2001-04) for developing partial 3D imaging algorithms.

\clearpage

{\small
\bibliographystyle{ieee_fullname}
\bibliography{e2eHS}
}

\end{document}